\documentclass[fleqn,usenatbib]{mnras}

\usepackage{newtxtext,newtxmath}

\usepackage[T1]{fontenc}

\DeclareRobustCommand{\VAN}[3]{#2}
\let\VANthebibliography\thebibliography
\def\thebibliography{\DeclareRobustCommand{\VAN}[3]{##3}\VANthebibliography}

\usepackage{graphicx}	
\usepackage{amsmath}	

\title[Orbital precession of S2 star in STVG]{Response to: Comment on "Orbital precession of the S2 star in scalar-tensor-vector gravity"}

\author[R. Della Monica et al.]{
Riccardo Della Monica,$^{1}$\thanks{E-mail: rdellamonica@usal.es}
Ivan de Martino,$^{1}$\thanks{E-mail: ivan.demartino@usal.es}
Mariafelicia de Laurentis,$^{2,3}$\thanks{E-mail: mariafelicia.delaurentis@unina.it}
\\
$^{1}$Universidad de Salamanca, Departamento de Fisica Fundamental, P. de la Merced, Salamanca, ES\\
$^{2}$Dipartimento di Fisica, Universit\'a
di Napoli {}``Federico II'', Compl. Univ. di
Monte S. Angelo, Edificio G, Via Cinthia, I-80126, Napoli, Italy\\
$^{3}$ INFN Sezione  di Napoli, Compl. Univ. di
Monte S. Angelo, Edificio G, Via Cinthia, I-80126, Napoli, Italy
}

\date{Accepted XXX. Received YYY; in original form ZZZ}

\pubyear{2023}

\begin{document}
\label{firstpage}
\pagerange{\pageref{firstpage}--\pageref{lastpage}}
\maketitle

\begin{abstract}
The explicit derivation for the orbital precession of the S2 star in the Galactic Center in the Scalar–Tensor–Vector Gravity is discussed and compared with previous research. The two different predictions are validated by numerically integrating the geodesic equations for a test particle.
\end{abstract}


Motivated by the recent advancement in the astrometric measurements of the proper motion of individual stars belonging to the S-stars cluster around the supermassive black hole (SMBH) Sagittarius A* (SgrA*) in the Galactic Center, in our original work, \citet{DellaMonica2022}, we 
computed stellar orbits around a SBMH in 
the Scalar-Tensor-Vector Gravity (STVG), an extension to the theory of General Relativity (GR) first proposed in \citet{Moffat2006}. Making use of the pre-pericenter data for the S2 star, recorded over more than two decades up to 2016 \citep{Gillessen2017}, and of the more recent detection of its post-Newtonian orbital precession \citep{Gravity2020}, for the first time we   
placed constraints on the deviations of STVG from GR on the very small scales (100-1000 AU) of the Galatic Center \citep{DellaMonica2022}. More recently, 
\citet{Turimov2022} discussed the calculations for the periastron precession of a test particle around a SMBH in STVG, 
that apparently contradict the relation for the orbital precession given in Eq. (16) of \citet{DellaMonica2022}. In this Response, we show how an erroneous assumption made in \citet{Turimov2022} leads to the seemingly contradictory results and how the two works can be reconciled.

\par

The STVG is a generally covariant alternative theory of gravity based on a modification of the Hilbert-Einstein action of GR. The theory introduces, in addition to the metric tensor field $g_{\mu\nu}$, a Proca-type massive vector field $\phi^\mu$ and elevates Newton's gravitational constant $G$ and the mass $\tilde{\mu}$ of the vector field to dynamical scalar fields that allow for an effective description of the variation of these “constants” with space and time.
The STVG field equations, obtained by minimization of the action, can be solved in vacuum assuming spherical symmetry. Further assumptions include that the measure of the gravitational coupling $G$ is constant, $\partial_\nu G = 0$, and has an enhanced (w.r.t. GR) value $G=G_N(1+\alpha)$ depending on a free dimensionless parameter $\alpha$, and that the mass of the vector field $\phi$, $\tilde{\mu}$, can be neglected when solving field equations for compact objects as black holes (BHs), because its effects manifest on kpc scales from the source. These assumptions lead to the expression of the space-time metric \citep{Moffat2015b}
\begin{align}
    ds^2=&\frac{\Delta}{r^2}dt^2-\frac{r^2}{\Delta}dr^2-r^2d\Omega^2\,,
    \label{eq:stvg-sch-metric}
\end{align}
where, assuming $c=1$ (the speed of light in vacuum),
\begin{align}
    &\Delta=r^2-2GMr+\alpha G_NGM^2,\label{eq:delta}\\
    &d\Omega^2=d\theta^2+\sin^2\theta d\phi^2,
\end{align}
which describes exactly the space-time around a point-like source with mass $M$. One of the key features of STVG is that the geodesic equations
possess a non-null right-hand side
\begin{align}
    \left(\frac{d^2x^\mu}{d\lambda^2}+\Gamma^\mu_{\nu\rho}\frac{dx^\nu}{d\lambda}\frac{dx^\rho}{d\lambda} \right)=\frac{q}{m}{B^{\mu}}_\nu\frac{dx^\nu}{d\lambda}.
    \label{eq:geodesic-equations}
\end{align}
due to the interaction of massive particles with the vector field $\phi^\mu$. As a matter of fact, $q$ acts as a coupling constant between the point-like particles of mass $m$ and the vector field $\phi^\mu$, resulting in an extra Lorentz-type force, usually called fifth force, which depends on the velocity of the particle. The constant $q$ is called the fifth-force charge of the particle and its sign is \emph{postulated} to be positive (implying a repulsive fifth force) in order to describe physically stable stars, galaxies, galaxy clusters, and agreement with solar system observational data \citep{Moffat2006b, Moffat2007, Moffat2009, Moffat2013, Moffat2014, Moffat2015a, Moffat2015b,DeMartino2017,DeMartino2020} and its value is assumed to be proportional to the mass of the particle itself, $q = \kappa m$, so that the weak equivalence principle can be recovered. The proportionality constant $\kappa$ is defined by,
\begin{equation}
    \kappa = \sqrt{\alpha G_N}.
\end{equation}
so that when $\alpha$ vanishes, both the metric and the geodesic equations in STVG reduce to their GR counterpart.

\par

Under the same spherical symmetry assumptions, the dynamical equations for the vector field $\phi^\mu$ reads 
\begin{align}
    &\nabla_\mu B^{\mu\nu} = 0,\\
    &\nabla_\sigma B_{\mu\nu} +\nabla_\mu B_{\nu\sigma}+\nabla_\nu B_{\sigma\mu} = 0,
\end{align}
where $\nabla_\mu$ is the covariant derivative operator related to the metric tensor and $B_{\mu\nu} = \nabla_\mu\phi_\nu-\nabla_\nu\phi_\mu$, can be solved, resulting in ({\emph{e.g.}, \citet{LopezArmengol2017})
\begin{equation}
    \phi_\mu = \left(-\frac{\sqrt{\alpha G_N}M}{r}, 0, 0, 0\right).
    \label{eq:vector_field}
\end{equation}
The vector field $\phi_{\mu}$ behaves like a purely electrical radial field generated by a point source located at the BH position, whose charge is $Q = \sqrt{\alpha G_N}M$ and whose action on massive test particles is a \emph{repulsive} force counteracting the enhanced gravitational constant $G$. Due to the presence of this vector field in the geodesic equations, the description of the motion of test particles  in terms of conserved quantities has to be modified accordingly. While the Lagrangian function per unit mass of the test particle can be defined in the usual way from the metric elements of Eq. \eqref{eq:stvg-sch-metric}\footnote{Here we report a typo in the sign of the normalization factor as written in \citet{DellaMonica2022}, that was correctly reported in \citet{Turimov2022}. However, as this was only a typographical error, in all the computations the correct sign was adopted.},
\begin{equation}
    2\mathcal{L} = g_{\mu\nu}\dot{x}^\mu \dot{x}^\nu = \left\{
    \begin{array}{ll}
        0\qquad&\textrm{(mass-less particle)}\\
        1\qquad&\textrm{(massive test particle)}
    \end{array}
    \right.
    \label{eq:lagrangian}
\end{equation}

one has to define generalized momenta, $\pi_\mu$, to incorporate the presence of the vector field. Due to the particular form of $\phi_\mu$ in Eq. \eqref{eq:vector_field}, only the time component of the 4-momentum is affected by the vector field \citep[see pages 898-900]{Gravitation2017}:
\begin{align}
    \pi_t &\equiv \frac{\partial \mathcal{L}}{\partial \dot{t}}-\frac{q}{m}\phi_t,\\
    \pi_r &\equiv \frac{\partial \mathcal{L}}{\partial \dot{r}},\\
    \pi_\theta &\equiv \frac{\partial \mathcal{L}}{\partial \dot{   \theta}},\\
    \pi_\phi &\equiv \frac{\partial \mathcal{L}}{\partial \dot{\phi}}.
\end{align}
Due to the symmetry of the Lagrangian with respect to the coordinates $t$ and $\phi$, the two quantities
\begin{align}
    E&\equiv \pi_t=\frac{\Delta}{r^2}\dot{t}+\frac{MG_N\alpha}{r}\label{eq:energy}\\
    L&\equiv -\pi_\phi=r^2\sin^2\theta\dot{\phi}\label{eq:angular_momentum}
\end{align}
are constants of motion and can be regarded as the specific energy and specific angular momentum of the particle at infinity (the term \emph{specific} refers to the fact that these quantities have to be regarded as energy and angular momentum per unit mass, \emph{i.e.} for $m=1$). Moreover, from
\begin{equation}
    \pi_\theta = -r^2\dot{\theta}\,,
\end{equation} and
\begin{equation}
    \frac{\partial\mathcal{L}}{\partial\theta} = -r^2\dot{\theta}\sin\theta\cos\theta,
\end{equation}
one can see that upon setting initial conditions so that the test particle initially lies on the equatorial plane $\theta = \pi/2$ with $\dot{\theta}=0$, the entire motion will take place on this plane and thus $\theta(\tau)=\pi/2$ identically. We can now use equations Eq. \eqref{eq:energy} and Eq. \eqref{eq:angular_momentum} to express $\dot{\phi}$ and $\dot{t}$ in terms of $E$ and $L$ and replacing them in  the equation Eq. \eqref{eq:lagrangian} for a test particle we can solve for $\dot{r}^2/\dot{\phi}^2=(dr/d\phi)^2$:
\begin{align}
    \left(\frac{dr}{d\phi}\right)^2 = \frac{r^2}{L^2}\left(E-\frac{M\alpha}{r}\right)^2-\Delta(r)\left(1+\frac{r^2}{L^2}\right),
    \label{eq:radial_motion}
\end{align}
where $G_N$ has been set to unity for simplicity. The latter equation is formally identical to Eqs. (5) and (6) in \citet{Turimov2022}, however, a significant difference is found, which is the different sign of the potential term related to the vector field in the energy definition. This can be clearly seen by comparing our Eq. \eqref{eq:energy} with Eq. (18) in \citet{Turimov2022}. The change in the sign of the vector field term has a major physical consequence. While we consider a repulsive vector field (\emph{i.e.} a fifth-force counteracting the increased gravitational attraction due to the increment in the gravitational coupling $G$) which is in accordance with Moffat's original idea \citep{Moffat2006} and with subsequent studies \citep{Moffat2006b, Moffat2007, Moffat2009, Moffat2013, Moffat2014, Moffat2015a, Moffat2015b, DeMartino2017,DeMartino2020}, in \citet{Turimov2022} an attractive field is erroneously considered, which is responsible for an even greater gravitational attraction generated by the central source on test particles. Starting with this erroneous assumption, the computations performed in \citet{Turimov2022} are thus not reproducing the expected results, leading to an excessively large rate of orbital precession, as one would expect from a stronger attraction between the test particle and the central source. In Appendix \ref{appendix:calculations} we report the correct detailed calculation of the orbital precession for the case of a repulsive fifth-force, which confirms our originally proposed formula for the first-order relativistic precession of a test particle:
\begin{equation}
    \Delta\phi_{\rm STVG} = \Delta \phi_{\rm GR}\left(1+\frac{5}{6}\alpha\right).
    \label{eq:precession_STVG}
\end{equation}

\begin{figure}
    \centering
    \includegraphics[width = \columnwidth]{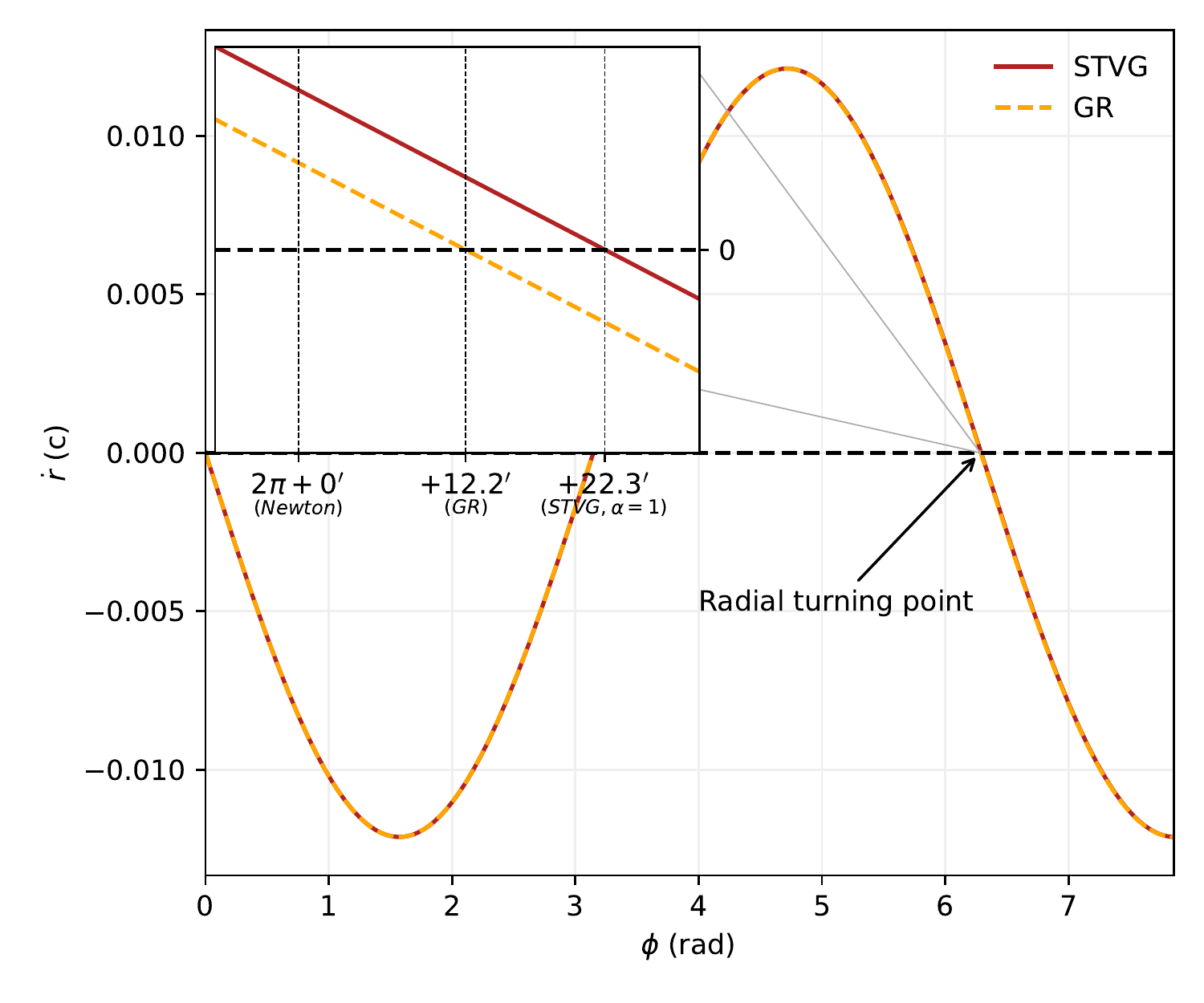}
    \caption{Numerical radial velocity profile of the S2 star for $\alpha = 0$ (dashed orange line) and for $\alpha = 1$ (solid red line) during one orbital period. The inset figure zooms into the apocentre passage (corresponding to the radial turning point $\dot{r}=0$). The GR orbit has a periastron shift of 12.2' per orbital period, while the STVG orbit has an enhanced value of the orbital precession of 22.3' per orbital period which is roughly 5/6 times greater than the GR case, in agreement with our analytical expression \eqref{eq:precession_STVG} for $\alpha = 1$.}
    \label{fig:precession_STVG}
\end{figure}

In order to double-check our results, we have computed a numerical integration of the geodesic equations \eqref{eq:geodesic-equations} for a time-like geodesic (whose orbital data correspond to those of the S2 star) for the case of a repulsive vector field $\phi^\mu$. For the sake of completeness, we have repeated the numerical computations also for the case of an attractive $\phi^\mu$, as erroneously considered in \citet{Turimov2022}. The orbital precession is quantified from the numerically integrated orbits as the difference from $2\pi$ of the angle spanned during one orbital period (\emph{i.e.} between two radial turning points). For this reason, in Figure \ref{fig:precession_STVG} we report the radial velocity profile of the S2 star as a function of the angular coordinate $\phi$. The plot illustrates both the $\alpha = 0$ (dashed orange line) case and the $\alpha = 1$ (solid red line) one, during one orbital period. Points in which $\dot{r} = 0$ correspond to radial turning points, \emph{i.e.} those points along the orbit where either a maximum or a minimum radial distance from the central source are reached (the apocentre and pericentre, respectively). As shown in the inset figure, which zooms into the apocentre passage, the GR orbit has gained a periastron shift of 12.2' after one orbital period, which is compatible with values reported in literature \citep{Gillessen2017, Gravity2020}. The STVG orbit (with the correctly assumed repulsive vector field $\phi^\mu$), on the other hand, is characterised by a greater value of the orbital precession of 22.3' per orbital period is observed. This is greater than the Schwarzschild precession in GR by a factor 5/6, which is in perfect agreement with our analytical expression \eqref{eq:precession_STVG} for $\alpha = 1$. In Figure \ref{fig:comparison}, on the other hand, we report the ratio to the GR precession of the numerically computed orbital precession of S2 for both the correct repulsive case for $\phi^\mu$ (black circles) and the erroneous attractive field (violet squares), as compared to the values predicted by the analytical formula Eq. \eqref{eq:precession_STVG}, originally presented in \citet{DellaMonica2022}, and the one reported in \citet{Turimov2022} for values of $\delta = 1$ and $\delta = 2$. The plot clearly shows the perfect agreement (in the range $\alpha \in [0,0.5]$) between the numerically computed values for the precession and our analytical expression \eqref{eq:precession_STVG} in the repulsive case (and, not surprisingly, agreement of the numerically computed precession in the attractive case with the $\delta = 2$ profile from \citet{Turimov2022}). 

\begin{figure}
    \centering
    \includegraphics[width = \columnwidth]{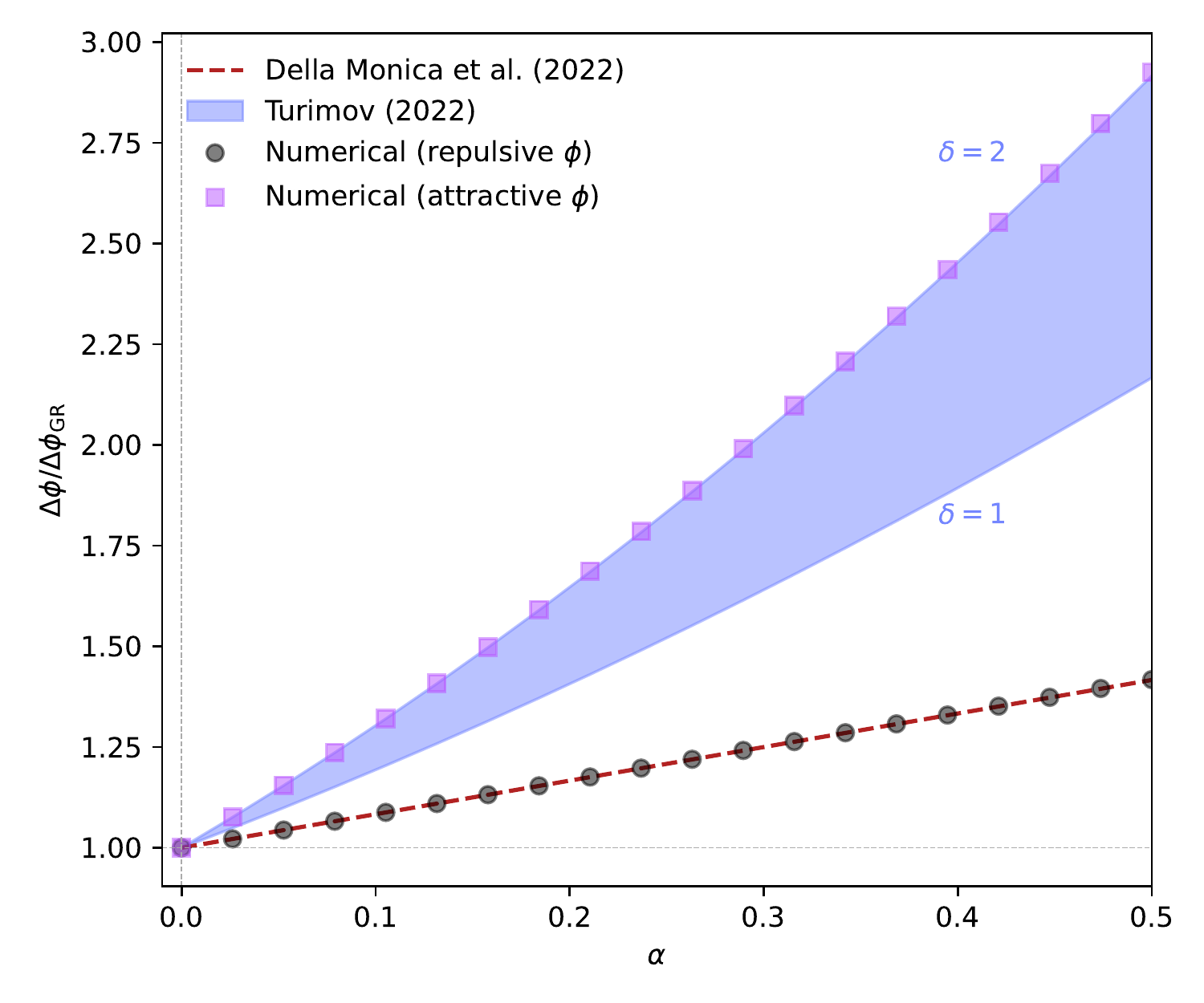}
    \caption{Ratio between the numerically computed orbital precession of S2 (for both the correct repulsive case for $\phi^\mu$ - black circles - and the erroneous repulsive field - violet squares) and the GR precession. We also report the values predicted by the analytical formula Eq. \eqref{eq:precession_STVG} (red dashed line) presented in \citet{DellaMonica2022}, and the one from \citet{Turimov2022} for values of $\delta = 1$ and $\delta = 2$ (solid blue lines). In the range $\alpha \in [0,0.5]$, the numerically computed values for the precession in the repulsive case agree with the prediction from the analytical expression. Moreover, the numerically computed precession in the attractive case agrees with the $\delta = 2$ profile from \citet{Turimov2022}).}
    \label{fig:comparison}
\end{figure}

The two results can be reconciled upon considering the correct signs in the energy definition in \citet{Turimov2022} which, on turn, would imply a different definition for the extra parameter $\delta$ thereby introduced. In this new definition, when particularized for the S2 star, one would find $\delta \ll 1$ (see Appendix \ref{appendix:calculations}) thus reducing the formula for the orbital precession found in \citet{Turimov2022} to our originally proposed expression.

\section*{Acknowledgements} RDM acknowledges support from Consejeria de Educación de la Junta de Castilla y León and from the Fondo Social Europeo. IDM acknowledges support from Ayuda IJCI2018-036198-I funded by MCIN/AEI/10.13039/501100011033 and: FSE “El FSE invierte en tu futuro” o “Financiado por la Unión Europea “NextGenerationEU”/PRTR. 
IDM is also supported by the project  PID2021-122938NB-I00 funded by the Spanish "Ministerio de Ciencia e Innovación" and FEDER “A way of making Europe", and by the project SA096P20 Junta de Castilla y León.

\section*{Data Availability Statement}
No new data were generated or analysed in support of this research.



\bibliographystyle{mnras}
\bibliography{biblio} 


\appendix

\section{Detailed calculations of the orbital precession}
\label{appendix:calculations}

In order to compute a first-order expression for the orbital precession for the S2 star in STVG (considering a repulsive vector field $\phi_\mu$), we apply the usual textbook calculations starting from the dynamical equation regulating the polar motion, $r(\phi)$, in the equatorial plane for a massive test particle, already given in Eq. \eqref{eq:radial_motion} of the main text, where for simplicity of notation we adopted $G=c=1$,
\begin{align}
    \left(\frac{dr}{d\phi}\right)^2 = \frac{r^4}{L^2}\left(E-\frac{M\alpha}{r}\right)^2-\Delta(r)\left(1+\frac{r^2}{L^2}\right).
\end{align}
Let's make the variable change $\xi = L^2/Mr$, so that the equation becomes
\begin{align}
    \left(\frac{L^2}{M\xi^2}\right)^2\left(\frac{d\xi}{d\phi}\right)^2 = \frac{L^6}{M^4\xi^4}\left(E-\frac{M^2\alpha}{L^2}\xi\right)-\Delta(r(\xi))\left(1+\frac{L^2}{M^2\xi^2}\right).   
\end{align}
If we now define the parameter $\sigma = (M/L)^2$, the latter equation can be rewritten as
\begin{align}
    \left(\frac{d\xi}{d\phi}\right)^2 = &\frac{1}{\sigma}(E-\alpha\sigma\xi)^2+\nonumber \\&-(1-2\sigma(1+\alpha)\xi+\sigma^2\alpha(1+\alpha)\xi^2)\left(\xi^2+\frac{1}{\sigma}\right)
\end{align}
where we have used the definition of $\Delta$ given in Eq. \eqref{eq:delta}. The latter equation can be differentiated with respect to $\xi$ resulting in
\begin{equation}
    \xi'' = 1 + \alpha(1-E)-\xi(1+\sigma\alpha)+3\sigma(1+\alpha)\xi^2-2\alpha(1+\alpha)\epsilon^2.
    \label{eq:xi_2}
\end{equation}
Where primes stand for derivatives with respect to the angle $\phi$. The non-homogeneous term $\delta=1-E$ appearing in this equation can be neglected for the purpose of this computation. In fact, if we restore all the dimensional constants and particularize it for the orbit of S2 (e.g. at pericenter), we obtain
\begin{align}
    \delta = 1-\frac{\Delta(r)}{r^2}\dot{t}-\frac{G_NM\alpha}{c^2r} &\sim 2.10\times10^{-5}\qquad\textrm{(for $\alpha = 0$)}\\
    &\sim 2.13\sim 10^{-5}\qquad\textrm{(for $\alpha = 2$)}
\end{align}
Neglecting the contribution of $\delta$, Eq. \eqref{eq:xi_2} becomes
\begin{align}
    \xi''+\xi(1+\sigma\alpha) = 1+3\sigma(1+\alpha)\xi^2-2\alpha(1+\alpha)\sigma^2.
    \label{eq:xi_3}
\end{align}
By applying a perturbative expansion of $\xi$ using $\sigma$ as a perturbation parameter (this is possible because, generally, $\sigma\ll 1$, e.g. for the S2 star $\sigma=1.88\times10^{-4}$), we can express the solution as
\begin{equation}
    \xi = \xi_0+\sigma\xi_1+\mathcal{O}(\sigma^2).
\end{equation}
If we substitute this solution in Eq. \eqref{eq:xi_3} stopping at first order, we get
\begin{align}
    &\xi''_0+\xi_0 = 1,\\
    &\xi''_1+\xi_1 = -\xi_0(\alpha-3(1+\alpha)\xi_0),
\end{align}
for the zero-th and first order, respectively. The zero-th order solution is the well known keplerian orbit, given by
\begin{equation}
    \xi_0(\phi) = 1+e\cos\phi
\end{equation}
where $e$ is the orbital eccentricity. Plugging this solution into the first order equation and neglecting all the subdominant contributions (\emph{i.e.} terms of higher order in the eccenetricity), we obtain:
\begin{equation}
    \xi''_1+\xi_1 \simeq 3+2\alpha + (5\alpha+6)e\cos\phi.
    \label{eq:xi_1}
\end{equation}
Due to the presence of the cosine at the right-hand side we look for solutions of the form $\xi_1(\phi) = A+B\phi\sin\phi$. Upon inserting this solution in Eq. \eqref{eq:xi_1} one easily finds:
\begin{align}
    A = 3+2\alpha\qquad;\qquad
    B = \frac{(5\alpha+6)e}{2}.
\end{align}
The full solution thus reads
\begin{equation}
    \xi(\phi) \simeq 1+e\cos\phi+(3+2\alpha)\sigma+\frac{(5\alpha+6)e}{2}\sigma\phi\sin\phi
\end{equation}
Using the perturbative expansion of the cosine function, i.e.
\begin{equation}
    \cos(\phi+\beta\phi) \simeq \cos(\phi)-\beta\phi\sin\phi+\mathcal{O}(\beta^2)
\end{equation}
with $\beta\ll 1$, one can rewrite the solution in the form
\begin{align}
    \xi(\phi) \simeq 1+(3+2\alpha)\sigma+\cos\left[\phi\left(1-3\sigma\left(1+\frac{5}{6}\alpha\right)\right)\right].
\end{align}
During one orbital period, thus, the orbit precedes by an angle
\begin{equation}
    \Delta\phi_{\rm STVG} = 6\pi\sigma\left(1+\frac{5}{6}\alpha\right)    
\end{equation}
and, remembering that the orbital precession predicted in GR is given by $\Delta\phi_{\rm GR} = 6\pi\sigma  = 6\pi G_NM/a(1-e^2)c^2$ (where $a$ is the semi-major axis of the orbit), we finally obtain the departure of the orbital precession in STVG from the one predicted by GR, given by
\begin{equation}
    \Delta\phi_{\rm STVG} = \Delta\phi_{\rm GR}\left(1+\frac{5}{6}\alpha\right)
\end{equation}

\bsp	
\label{lastpage}
\end{document}